\documentclass{article}
\usepackage{amsmath,amsbsy,amsfonts,amssymb}
\usepackage{subfigure}
\usepackage{graphicx}
\usepackage{hyperref}

\usepackage[left=1.5cm,right=1.5cm,top=2cm,bottom=2cm]{geometry}

\author{Irving Rond\'on and F.~ Soto-Eguibar,\\
 Instituto Nacional de Astrof\'isica \'Optica y Electr\'onica,\\ Puebla, C.P. 72840 M\'exico.
}

\title{Generalized optical theorem for propagation invariant beams}
\begin{document}

\maketitle
\begin{abstract}
	Many practical applications require the analysis of electromagnetic scattering properties of local structures   using  different  sources of illumination. The Optical Theorem (OT) is a useful  result  in scattering theory, relating the extinction  of a structure  to the scattering amplitude in the	forward direction. The  most common derivation of the OT is given for plane waves but  advances in optical engineering now allow  laser beam shaping, which might require an  extended theorem where the impinging source is a structured field. In this work, we derive an expression for the optical theorem based on classical electromagnetic theory, for probe sources given in terms of propagation invariant beams. We obtain a general expression for the differential scattering  cross section  using the integral scattering amplitude approximation in the far field. We also analyze the scattering problem of a zero order  Bessel beam by a dielectric sphere, under the Rayleigh approximation by varying the angle of incidence.
\end{abstract}

\section{Introduction}
This work is focused  on a fundamental relation in scattering theory, the so-called optical cross-section theorem or simply the Optical Theorem (OT), which describes the rate at which energy is distributed  from a probing incident wave field  by a scattering object, due to re-radiation and absorption by the scatterer \cite{Jackson}. The optical theorem has a long and  interesting  history; it appeared in electromagnetic theory  more than one hundred years ago and similar theorems can be found in acoustical scattering and quantum mechanics \cite{RNewton}. This classical problem is covered in advanced electromagnetic books where it is always presented for plane waves \cite{Bohren}. Essentially, the optical theorem states that the rate at which energy is extincted due to scattering and absorption at the scatterer is proportional to the imaginary part of the forward scattering amplitude, corresponding to the direction of propagation of the incident field \cite{Tsang2000,Ishimaru1991,Berg2008,Berg2008a}. The OT  provides an analytic tool to determine some of the objects physical properties (i.e., shape, size, concentration, density, absorption, conductivity, etc.), from the scattered field.\\
Previous works have attempted to unify the OT. An example of those is the generalized optical theorem for scalar fields \cite{Carney1}, which is found  in the near-field optics \cite{Carney2}. In this context, the problem is solved for a non homogeneous Helmholtz equation,  $\nabla^2 \psi + k^2 \psi = -4 \pi k^2 \eta \psi$ being $k = \omega/c$, $\eta$ the dielectric susceptibility, and  $\psi$ the scattering potential solution of the  partial differential equation. A version of the generalized  optical theorem has been discussed in the  scalar and  vector electromagnetic approaches in a wide context of applications \cite{Kees}. An extended formalism has been presented in the field of acoustics \cite{Zhang2013}. The solution of the  scattering problem in quantum mechanics for  non-plane waves\cite{Gouesbet2009}, for anisotropic media \cite{Marengo2013}, and  for arbitrary scalar and vectorial fields in cylindrical coordinates \cite{Mitri2016,Mitri2015} have been also analyzed. Recently, Marengo and Tu provided an OT that describes the energy budget of wave 	scattering phenomena in time domain \cite{Marengo1} and its applications in transmission lines \cite{Marengo2}. \\
However, it is not the aim of this manuscript to list the full range of applications and references where the OT has been used. Instead, we aim to present an explicit derivation of a complete general expression of the OT based on Maxwell equations that can be applied to any invariant beam. To the best of our knowledge, this has not been presented before. 
An invariant beam, also known as a \textquotedblleft non-diffracting beam\textquotedblright, propagates indefinitely without changes in its transverse intensity distribution \cite{Bajer1996a, Bajer1997}. These beams can also be represented as an infinite superposition of plane waves \cite{NietoVesperinas1991}. These optical fields are solution of the transverse Helmholtz wave equation $\nabla^2_T \varphi + k_T^2 \varphi = 0 $ in  Cartesian, circular cylindrical, parabolic cylindrical, and elliptical cylindrical coordinates, where $k_T$ is the transversal wave vector; the solutions of this  equation are the well known plane, Bessel \cite{Durnin1987}, Mathieu \cite{GutierrezVega2000}  and Weber beams \cite{Bandres2004}, respectively. The use of  such fields  in the experimental and theoretical works has attracted considerable attention \cite{Hugo2014}, from quantum mechanics \cite{Rocio,BM1,BM2,BM3} to space communications \cite{Willner}. New scenarios have been opened for scattering problems \cite{Fred,QZhan,Alua,Geffrin}; however, none of them explores the possibility of using these fields on any of the new technological challenges that range from nano and micro-photonics to science and engineering of antennas, metamaterials and electromagnetic devices, among others.\\
The main goal of this research is to  present  a derivation  of the  general OT  based on Maxwell equations and the standard literature for propagation invariant beams, where the plane wave case is a particular case, using the well-known far field approximation \cite{Mischenko2006a,Mishchenko2006b}. In order to illustrate our results we revisit the classical and standard scattering  elastic problem of a dielectric sphere in the Rayleigh regime, for which the incident field  can be  any propagation  invariant beam.   The  problem becomes clearer studying the case when the electric incident field is a a zero order Bessel beam, due to its wide range of applications \cite{McGloin,CLMariscal}. Finally, we present some conclusions and possible applications.

\section{Optical theorem}
\begin{figure}[htbp!]
	\centering
	\includegraphics[width=9 cm]{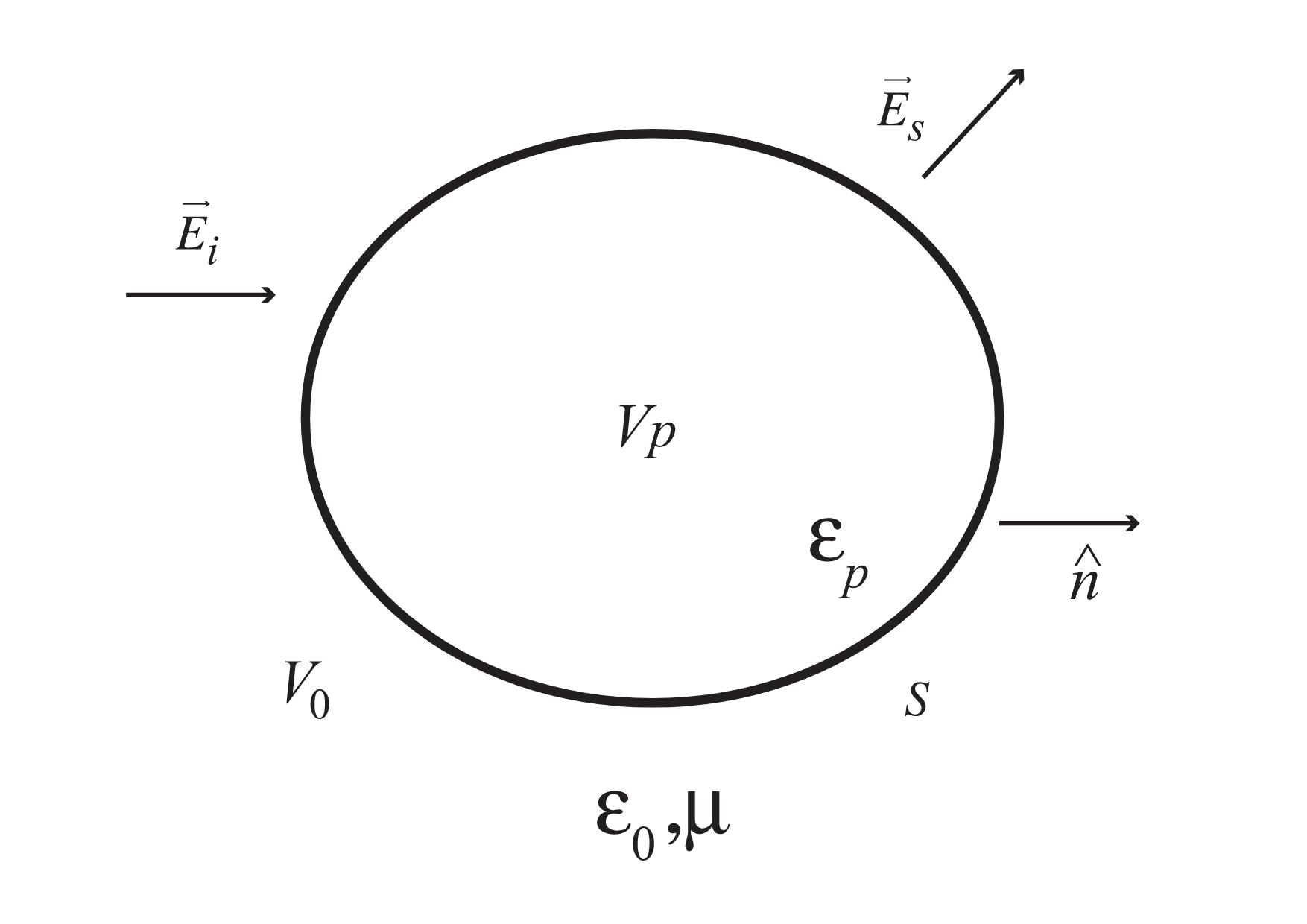}
	\caption{Geometrical configuration for the generalization of the optical theorem.}
	\label{figura1}
\end{figure}

\begin{figure}
	\centering
	\includegraphics[width=9 cm]{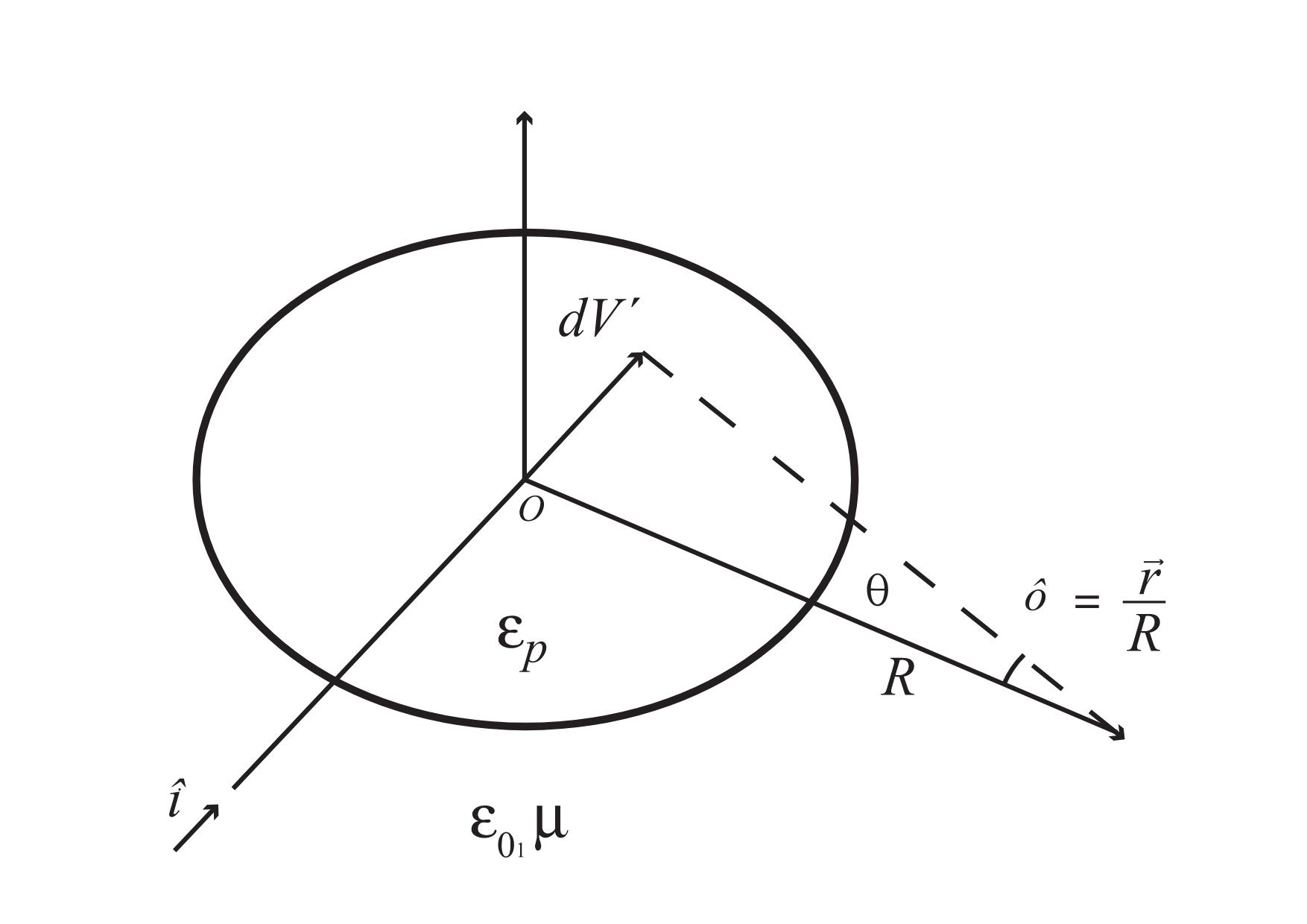}
	\caption{Diagram of an incident field upon an object in the far field approximation.}	
	\label{figura2}
\end{figure}

Let us consider a scatterer particle of arbitrary form and size, volume $V$, and a complex permittivity 
\begin{equation}
\varepsilon_p\left( \vec{r}\right) =\varepsilon_0 \varepsilon_r\left( \vec{r}\right) =\varepsilon_0\left[  \varepsilon_r'(\vec{r})+i \varepsilon_r''(\vec{r})\right],	
\end{equation} 
where $\varepsilon_0$ is the vacuum permittivity. The medium surrounding the scatterer particle is lossless and its permittivity $\varepsilon$ is real. We assume also that the scatterer particle and the surrounding medium are non magnetic, so that they have  magnetic permeability $\mu=1$, see Figure \ref{figura1}. The fields $\vec{E}_i(\vec{r}),\, \vec{H}_i(\vec{r})$ correspond to an invariant beam that strikes the scatterer particle;  the scattered electromagnetic field is represented by $\vec{E}_s(\vec{r}),\, \vec{H}_s(\vec{r})$ and $\vec{E}_p(\vec{r}),\, \vec{H}_p(\vec{r})$ describe the electromagnetic field inside the scatterer particle, as shown in Figure \ref{figura1}.
The time-averaged power absorbed by the scatterer object is given by \cite{Ishimaru1991,Tsang2000}
\begin{equation}
P_a=\frac{\varepsilon_0}{2}\int_{V}\omega \, \varepsilon''_r\left(\vec{r} \right)  \, \left| \vec{E}_p \left( \vec{r} \right) \right|^2   dV.
\end{equation}
If the Poynting vector of the scattered field is given by the following expression 
\begin{equation}
\vec{S}_s=\frac{1}{2}\mathbf{Re}\left(\vec{E}_s \times \vec{H}_s^* \right), 
\end{equation}
then the time-average scattered power is \cite{Tsang2000}
\begin{equation}
P_s=\int_{S_{\infty}} \vec{S}_s \cdot \hat{n} \, da
=\int_{\mathrm{sup}} \vec{S}_s \cdot \hat{n} \, da,
\end{equation}
where $\mathrm{sup}$ is the surface around the volume $V$ of the scatterer particle and $S_\infty$ is the surface at infinity. Assuming nothing else that the correctness of the Maxwell equations, it is possible to show that \cite{Tsang2000}
\begin{equation}\label{ecPaTotal}
P_{a} + P_{s}=  -\int_{\mathrm{sup}}
\vec{S}'\cdot \hat{n}    da,
\end{equation}
with
\begin{equation}
\vec{S}'=\frac{1}{2}\mathbf{Re} \left(  \vec{E}_i \times \vec{H}_s^*  +\vec{E}_s \times \vec{H}_i^*\right).
\end{equation}
Bajer and Horak \cite{Bajer1996a} have shown that for an invariant beam $\nabla \cdot \left( \vec{S}_i +\vec{S}_s \right)  =0$. Thus, using the divergence theorem, it is easy to prove that \cite{Bajer1996a, Bajer1997}
\begin{equation}
\int_{\mathrm{sup}} \mathbf{Re}\left( \vec{E}_i \times \vec{H}_s^*\right) \cdot \hat{n}  \,  da =
\int_{\mathrm{sup}} \mathbf{Re}\left( \vec{E}_i^* \times \vec{H}\right) \cdot \hat{n}  \,  da,
\end{equation}
and that
\begin{equation}
\int_{\mathrm{sup}} \mathbf{Re}\left( \vec{E}_s \times \vec{H}_i^*\right) \cdot \hat{n}  \,  da =
\int_{\mathrm{sup}} \mathbf{Re}\left( \vec{E} \times \vec{H}_i^*\right) \cdot \hat{n}  \,  da,
\end{equation}
where $\vec{E}$ y $\vec{H}$ denote the total fields; i.e., $\vec{E}=\vec{E}_i+\vec{E}_s$ and $\vec{H}=\vec{H}_i+\vec{H}_s$. Substituting these last expressions in (\ref{ecPaTotal}), using the divergence theorem and expanding the divergence of the cross product obtained, we arrive to the following expression for the total power absorbed and scattered,
\begin{equation}
\label{ecPaPs}
\begin{split}
P_{a} + P_{s}
&=  -\frac{1}{2} \mathbf{Re} \int_{V}
\left[\left(\nabla\times\vec{E}_i^* \right)\cdot \vec{H}  - \vec{E}_i^* \cdot\left(\nabla\times\vec{H} \right)  
\right.   \\  &  \left.  
\qquad \qquad \qquad + \left(\nabla\times\vec{E} \right)\cdot \vec{H}_i ^*  -    \vec{E}\cdot \left(\nabla\times\vec{H}_i^* \right)\right]     dV.
\end{split}
\end{equation}
If we substitute the Maxwell equations,
\begin{align}
\label{ecMax1}
\nabla \times \vec{E} &=  i \omega \mu \vec{H}, \\
\label{ecMax2}
\nabla \times \vec{H} &= - i \omega \varepsilon_0 \vec{E},
\end{align}
in \eqref{ecPaPs}, we can recover,
\begin{equation}
\label{ec:ImPaPs}
\begin{split}
P_{a} + P_{s} & =  -\frac{1}{2} \mathbf{Re} \int_{V}
\left[\left( -i \omega \mu \vec{H}_i^* \right)\cdot \vec{H}  - \vec{E}_i^* \cdot\left( - i \omega \varepsilon_p \vec{E} \right) \right.     \\  &  \left.  
\qquad \qquad \qquad + \left(  i \omega \mu \vec{H} \right)\cdot  \vec{H}_i ^* - \vec{E}\cdot \left(  i \omega \varepsilon_0 \vec{E}_i^* \right)\right] dV   \\
& =  \frac{1}{2} \, \mathbf{Im} \int_{V}
\omega \left( \varepsilon_p - \varepsilon_0\right) \vec{E}_i^*\cdot\vec{E} \,     dV,
\end{split}
\end{equation}
where, for simplicity, we have avoided  writing the $\vec{r}$-dependence of the permittivity inside the particle. Equation  \eqref{ecMax2} can be written as
\begin{equation}
\label{ecMaxwellJeq}
\nabla \times \vec{H} = - i \varepsilon_0 \omega \vec{E} + \vec{J}_{\mathrm{eq}},
\end{equation}
where we have defined the current density 
\begin{align}
\label{ecJ}
\vec{J}_{\mathrm{eq}} (\vec{r}) =\begin{cases}  -i\omega \varepsilon_0  \left[ \varepsilon_r (\vec{r}) - 1 \right]\vec{E},&  \text{inside}\,\,  V, \\ 
0, & \text{outside}\,\, V,
\end{cases}
\end{align}
which can be thought as the producer of  the scattered field. We introduce now the Hertz potential
\begin{equation}\label{ecPotH}
\vec{\Pi}_s (\vec{r}) = \frac{1}{ - i \omega \varepsilon_0 } \int_V G_0 (\vec{r},\vec{r}') J_{\mathrm{eq}}(\vec{r}') dV',
\end{equation}
where the Green function for the Hemholtz equation is
\begin{equation}\label{ecFuncGreen}
G_0 (\vec{r},\vec{r}')=   \frac{   e^{ ik \vert \vec{r} - \vec{r'} \vert}}{4\pi \vert \vec{r} - \vec{r'} \vert },
\end{equation}
which represents the scattered field. If we substitute \eqref{ecJ} in \eqref{ecPotH}, we obtain
\begin{align}\label{ecPotHert1}
\vec{\Pi}_s (\vec{r}) =  \int_V
\left[ \varepsilon_r(\vec{r'}) - 1 \right]  \vec{E}(\vec{r'}) G_0 (\vec{r},\vec{r'})  dV'.
\end{align}
In terms of the Hertz potential the scattered electric and magnetic fields can be written as \cite{Ishimaru1991}
\begin{align}
\vec{E}_s (\vec{r}) & = \nabla \times \nabla \times  \vec{\Pi}_s (\vec{r}),   \label{ecEdisp} \\ 
\vec{H}_s  (\vec{r}) & = - i\omega \varepsilon \nabla \times  \vec{\Pi}_s (\vec{r}).  \label{ecHdisp}
\end{align}
In the far field approximation, the quantity $\frac{1}{\vert \vec{r} - \vec{r'} \vert}$ can be approximated by $\frac{1}{R}$, where $R=\left| \vec{r} \right| $. However, in the case of $ik\vert \vec{r} - \vec{r'} \vert$,
the correct far field approximation is \cite{Mischenko2006a,Mishchenko2006b} 
\begin{equation}\label{ecExpBino}
\vert \vec{r} - \vec{r'} \vert  = \sqrt{R^2 -2 R \vec{r'}\cdot \hat{o} + \vec{r}^2}  \approx R - \vec{r'} \cdot \hat{o},
\end{equation}
where $\hat{o}=\frac{\vec{r}}{R}$ is the unit vector directed to the observation point. As shown in Figure \ref{figura2}, the Green function \eqref{ecFuncGreen} takes the form,
\begin{equation}\label{ecFGreenCampLej}
G_0 (\vec{r},\vec{r'})=   \frac{e^{ i k R - ik \hat{r'}\cdot \hat{o}}}{4\pi R }.
\end{equation}
It is straightforward to show that
\begin{equation}
\nabla  \left( \frac{e^{ i k R}}{R}\right)\approx  ik\hat{o} \frac{e^{ i k R}}{ 4 \pi R}.
\end{equation}
Thus, the scattered field, Equation \eqref{ecEdisp}, can be written as
\begin{equation}
\vec{E}_s (\vec{r})  =    \frac{e^{i k R}}{4 \pi R} k^2  \int_V  \Big[ \vec{E}  - \hat{o}\left( \hat{o}\cdot \vec{E}\right) \Big]\left[ \varepsilon_r(\vec{r'})-1\right] e^{-i k \vec{r'}\cdot \hat{o}}  dV'.
\end{equation}
The previous equation can be written as
\begin{equation}
\label{ec:OndaEsfD}
\vec{E}_s (\vec{r}) = \frac{e^{i k R}}{ R} \vec{F}(\hat{i}, \hat{o}),
\end{equation}
where we have defined the scattering amplitude
\begin{equation}\label{ecfunForm}
\vec{F}(\hat{i}, \hat{o})  = \frac{k^2}{4 \pi}\int_V  \left[ \vec{E} - \hat{o}(\hat{o}\cdot \vec{E} ) \right] \left[ \varepsilon_r(\vec{r}')-1\right] e^{-i k \vec{r'}\cdot \hat{o}}  dV',
\end{equation}
and where the unit vector $\hat{i}$ is the direction of the impinging electric field; physically, equations \eqref{ec:OndaEsfD} and \eqref{ecfunForm} represent how a spherical wave is modulated. The scattering amplitude expresses  how the particle responds to the incident field and where that field is observed after the interaction with the particle.\\ 
We make now $\hat{o}=\hat{i}$ and we get the forward scattering amplitude
\begin{equation}
\label{FunAmplitude}
\vec{F}(\hat{i},\hat{i}) =
\frac{k^2}{4 \pi } \int _V   \vec{E} (\vec{r}')
\left[\varepsilon _r (\vec{r}')-1\right] 
e^{ - i k \vec{r}'\cdot \hat{i}}  dV'.
\end{equation}
We write the propagation invariant incident beam in its plane-wave representation
\cite{Whitaker,GutierrezVega2000,Bandres2004,NietoVesperinas1991}
\begin{align}
\label{ec:EoInv}
\vec{E}_i(\vec{r})=
\int_{-\pi}^{\pi}    \hat{e}_i
A (\phi )  e^{ - i k \hat{i}  \cdot \vec{r} }  d\phi,
\end{align}
where $\hat{e}_i$ is the polarization of the incident field, $k$ is the wave vector  
that can be related to the cone angle defined for invariant beams \cite{Durnin1987}; the function $A(\phi)$ represents the relative amplitude of the plane wave superposition \cite{NietoVesperinas1991}. Multiplying Eq. \eqref{FunAmplitude} by $A\left(\phi \right)$ and integrating it with respect to $\phi$ from $-\pi$ to $+\pi$, we get
\begin{align}
\int_{-\pi}^{\pi} 
\vec{F}(\hat{i},\hat{i})\cdot \hat{e}_i A (\phi)  d\phi  &=
\frac{k^2}{4 \pi } \int_{-\pi}^{\pi}  d\phi \int _V  \vec{E} (\vec{r}')
\left[\varepsilon _r (\vec{r}')-1\right] 
\cdot \hat{e}_i A (\phi ) e^{  - i k  \vec{r}'\cdot \hat{i}} dV'
\nonumber \\ &
= \frac{k^2}{4 \pi }  \int _V   \vec{E} (\vec{r}')
\left[\varepsilon _r (\vec{r}')-1\right] dV'
\cdot \int_{-\pi}^{\pi}   \, \hat{e}_i A (\phi ) e^{  - i k   \vec{r}'\cdot \hat{i}} d\phi; 
\end{align}
thus, using the representation \eqref{ec:EoInv} for an invariant incident field,
\begin{align}
\label{ec:InvPaPs}
\vec{F}(\hat{i},\hat{i})\cdot  \hat{e}_i \int_{-\pi}^{\pi}  
A (\phi)  d\phi  =
\frac{k^2}{4 \pi } \int _V   
\left[\varepsilon _r (\vec{r}')-1\right]  \vec{E} (\vec{r}') \cdot   \vec{E}_i(\vec{r}') dV'.
\end{align}
As $\int_{-\pi}^{\pi} A(\phi)  d\phi$ is a constant, equating integrals \eqref{ec:ImPaPs} and \eqref{ec:InvPaPs} gives us 
\begin{equation}
P_a +  P_s = \frac{2\pi} {k \eta} 
\mathbf{Im} \, \left[ 
\vec{F}(\hat{i},\hat{i})\cdot \hat{e}_i\right],
\end{equation}
which is  the  well known  optical theorem \cite{Jackson,RNewton,Bohren,Tsang2000,Ishimaru1991}. The results  presented so far are valid for any incident beam which can be written in the form given in \eqref{ec:EoInv}.  \\
Using the definition of the extinction transversal section as
$ \sigma_{ext} =  \sigma_{a} + \sigma_{s}$, where $\sigma_{a}$ is the absorbed transversal section, and $ \sigma_{s}$  is the scattering transversal section \cite{Bohren}, we can write the optical theorem as 
\begin{equation}
\label{ec:SecTO}
\sigma_{ext} = \frac{4\pi} {k \vert \langle\vec{S}_i \rangle\vert} 
\mathbf{Im} \, \left[ \vec{F}(\hat{i},\hat{i})\cdot \hat{e}_i \right],
\end{equation}
where $ \vert \langle\vec{S}_i \rangle\vert$ is the averaged  incident Poynting vector. 

\section{Example for  Rayleigh scattering}
In this section, the scattering of an incident propagation invariant beam by a dielectric sphere with constant relative permittivity $\varepsilon_r$ and radius $a$ is presented \cite{Ishimaru1991,Jackson,Bohren,Tsang2000}. It is well known that the electric field $\vec{E}$ inside a dielectric sphere  immersed in an electric field $ \vec{E}_i $ is given by \cite{Jackson,Ishimaru1991}
\begin{equation}
\label{EcCampEsfDiel}
\vec{E}= \frac{3}{2 + \varepsilon_r} \vec{E}_i.
\end{equation}
As the impinging field is an invariant beam, we substitute its plane wave representation, given by Eq. \eqref{ec:EoInv}, into \eqref{EcCampEsfDiel} and the result in Eq. \eqref{ecfunForm}. Then, under  the Rayleigh approximation $ e^{-i k \vec{r}'\cdot \hat{o}} \approx 1$ \cite{Fraser,Young}, the integral over the volume of the sphere can be easily calculated  to yield  
\begin{equation}
\label{ec:Finvariant}
F(\hat{i}, \hat{o})   =   k^2 a^3  \frac{ \varepsilon_r - 1 }{\varepsilon_r +2 }    
\left[ \hat{e}_i -  \left( \hat{o}\cdot  \hat{e}_i   \right)\hat{o} \right]
\int_{-\pi}^{\pi} d\phi  A (\phi ) e^{ - i k \hat{i}  \cdot \vec{r} },    
\end{equation}
where $k$ is the wave vector  and $a$ is the sphere radius. Using  \eqref{ec:Finvariant} as the differential scattering cross section  $\sigma_{\text{d}} \equiv \vert F(\hat{i}, \hat{o}) \vert^2$ and the appropriated relative amplitude of the plane wave superposition $A\left(\phi\right) $,  the  differential scattering cross section for any invariant beam can be obtained. Integration of this equation over the solid angle $d\Omega$ gives the scattering cross section \cite{Ishimaru1991}; the polarization as a function of the scattered radiation can be derived as well \cite{Jackson}. It is important to note that when the angular modulation function $A(\phi) $ is a  delta function, and the incident beam is a plane wave, the classical solution for Rayleigh scattering  is recovered \cite{Jackson}.\\
Using these results, we are also able to  study the case when the incident  field is a Bessel beam.  These beams were introduced  by Durnin \cite{Durnin1987}, and have attracted considerable attention  due to their properties of transverse propagation invariance and self-reconstruction, among others \cite{Hugo2014}. In addition, it is well know that Bessel beams carry both linear and angular momentum that can be transferred to atoms, molecules and particles \cite{McGloin, CLMariscal}.  For this case $A(\phi)=e^{im\phi}$ \cite{Zhang2013}, and after  substitution into \eqref{ec:Finvariant}, we obtain
\begin{equation}
\label{ec:AmplBessel}
F(\hat{i}, \hat{o})   =   k^2 a^3  \frac{ \varepsilon_r - 1 }{\varepsilon_r +2 }    
\left[ \hat{e}_i -  \left( \hat{o}\cdot  \hat{e}_i   \right)\hat{o} \right]
J_{m}(k_{T} \rho) e^{i m \phi},
\end{equation}
where $\rho= \sqrt{x^2+ y^2}$ is the transverse radius, $m$  is an integer, $k_{T}= k \sin \beta$ is the transversal wave vector, and  $\beta$ is the value of the half-cone angle \cite{Durnin1987}; the expression \eqref{ec:AmplBessel} is the scattering field amplitude for any Bessel beam of $m$-th order. It is important to remark that in \cite{Marston1,Marston2}, the authors have used the partial wave series in the far field approximation to calculate a form function equivalent to equation \eqref{ec:AmplBessel}, for the scattering of an acoustic helicoidal Bessel beam by a sphere centered on the axis; they  also calculated the acoustic radiation force exerted by the beam on the sphere in an inviscid fluid.  Remarkably, using the forward scattering amplitude, as presented in  \cite{Shimuzu,Gordon,Drake} for the Mie regime, allows  this approach to be extended due to the fact that any invariant beam can be constructed  as a plane wave superposition.
For our case, the differential scattering cross section is given by 
\begin{equation}
\label{ec:dScatBessel}
\sigma_{\text{d}}(\theta)  =  k^4 a^6 \left( \frac{ \varepsilon_r - 1 }{\varepsilon_r +2 } \right)^2   J_{m}^2(k_{T} \rho)  \left[ (1 + \cos^2 \theta)/2  \right],
\end{equation}
where we have used the standard  scattering geometry  \cite{Shimuzu,Gordon,Drake} to relate the scattering angle  $\theta = \arccos ( \hat{i}\cdot \hat{o} ) $ to the unit vectors  $\hat{i}$  and $\hat{o}$ at  a particular point \textit{P},  where the scattered radiation is observed. On the other hand,   as in the case of a plane wave, if the incident field is unpolarized the differential scattering function is the average  over parallel and perpendicular incidents fields. This gives   $\sigma_{\text{d}}(\theta) = \frac{1}{2} \left[  \sigma^{\perp}_{\text{d}}(\theta) +\sigma^{\parallel}_{\text{d}}(\theta) \right]$ and  $\hat{o}\cdot \hat{e}_i=0$, if $\hat{e}_i$ is perpendicular to the scattering plane, and $\hat{o}\cdot \hat{e}_1= \sin\theta$, if $\hat{e}_i$ lies in the plane. Using this information we can calculate the polarization scattering function \cite{Jackson}
\begin{equation}
\label{ec:fscatBessel}
\Pi(\theta) =  \frac{ \sigma^{\perp}_{\text{d}}(\theta)-\sigma^{\parallel}_{\text{d}}(\theta)  }{ \sigma^{\perp}_{\text{d}}(\theta)+\sigma^{\parallel}_{\text{d}}(\theta) }=\frac{  \sin^2\theta  }{1+ \cos^2 \theta} J_{m}^2 \left( k_T \rho\right).
\end{equation}
By integrating the equation \eqref{ec:dScatBessel} over $d\Omega$, it is straightforward to obtain   $\sigma_{\text{s}}$,  the scattering transversal section, as
\begin{align}
\label{ec:DiffScattering}
\sigma_{\text{s}} (\theta)= \int_{0}^{\pi} \int_{0}^{2\pi} \sigma_{\text{d}}(\theta)
d\Omega  =  \frac{8 }{3} (\pi a^2)   k^4 a^4 \left(\frac{ \varepsilon - \varepsilon_0}{\varepsilon + 2\varepsilon_0 }\right)^2 J_{m}^2 \left( k_T \rho\right),
\end{align}
while the absorbed transversal section $\sigma_\text{a}$ has to be obtained from the Poynting vector \cite{Jackson,Bohren,Tsang2000,Ishimaru1991}, since  it would result  in $\sigma_{\text{ext}}=0$ otherwise. Considering a transparent sphere (non-absorbing medium)  for which 
$ \sigma_{\text{ext}} =  \sigma_{\text{a}} + \sigma_\text{s}$, (meaning that $\sigma_{a} \approx 0$ \cite{Bohren}),  and using  equation \eqref{ec:DiffScattering} the scattering extinction efficiency  yields
\begin{align}
\label{ec:ScatterEff}
Q_{\text{ext}}=  \frac{\sigma_{\text{s}} (\theta)}{A_\text{t}}=  \frac{8 }{3}    (ka)^4 \left(\frac{ \varepsilon - \varepsilon_0}{\varepsilon + 2\varepsilon_0 }\right)^2 J_{m}^2 \left( k_T \rho\right),
\end{align}
where $A_{t}$ is the transversal area. This equation can be conveniently written as
\begin{align}
\label{ec:Qext}
Q_{\text{ext}}=   \frac{8 }{3}    x^4 \left(\frac{ n^2  - 1 }{  n^2 + 2}\right)^2 J_{m}^2 \left( k_T \rho\right),
\end{align}
where  $x= k a$ is a scattering parameter factor related to the wavelength and the sphere radius \cite{Bohren}, and we have represented the index of refraction as $ n^2 \equiv  \varepsilon/\varepsilon_0 $. Note that in the limiting case $m=0$ and $\rho \rightarrow 0 $ , $ J_{m} \left( k_T \rho\right) \rightarrow 1$; thus, this expression physically behaves as a plane wave and the Rayleigh scattering law is recovered. \\
\begin{figure}
	\centering
	\includegraphics[scale=0.5]{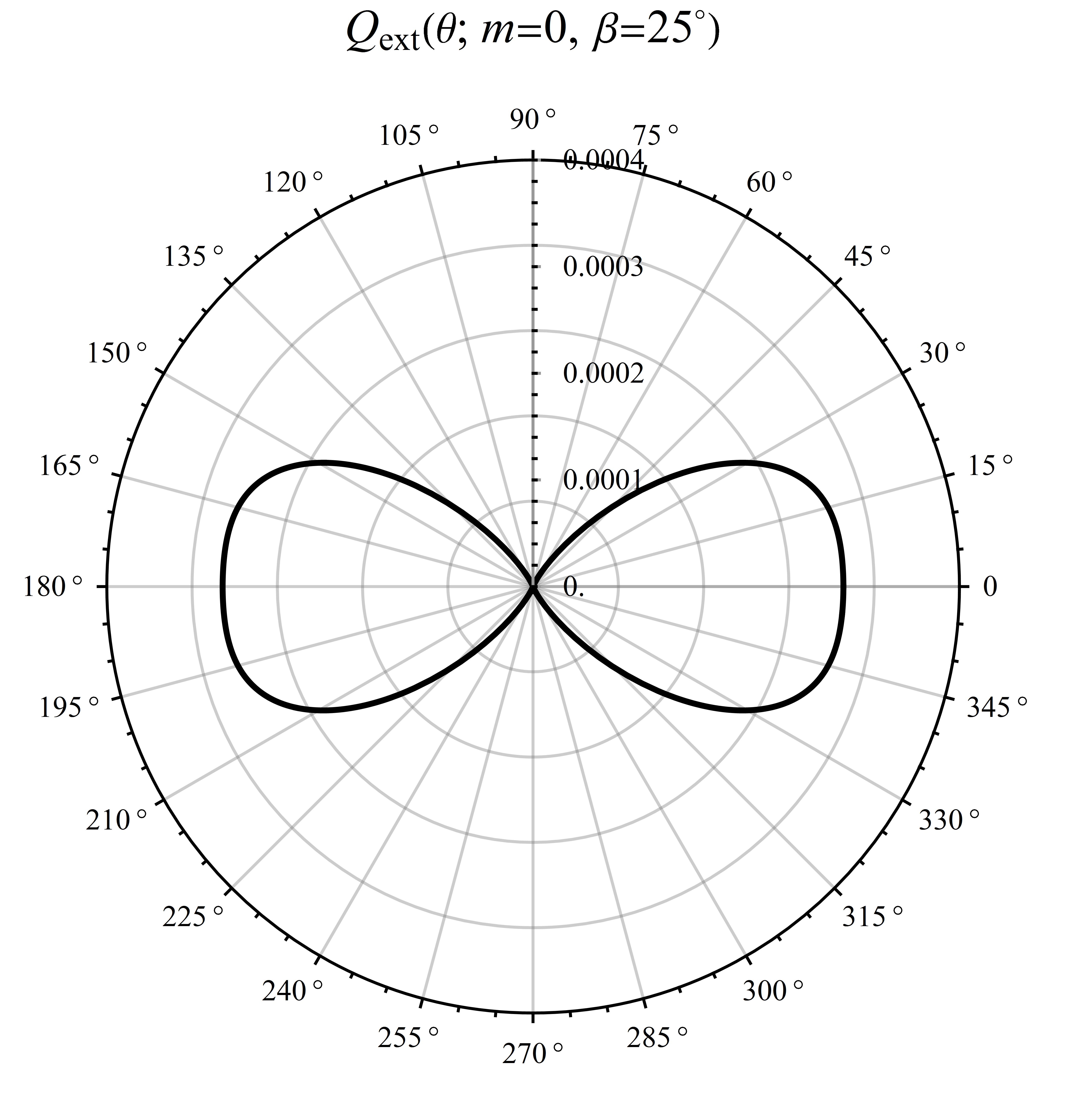}
	\includegraphics[scale=0.5]{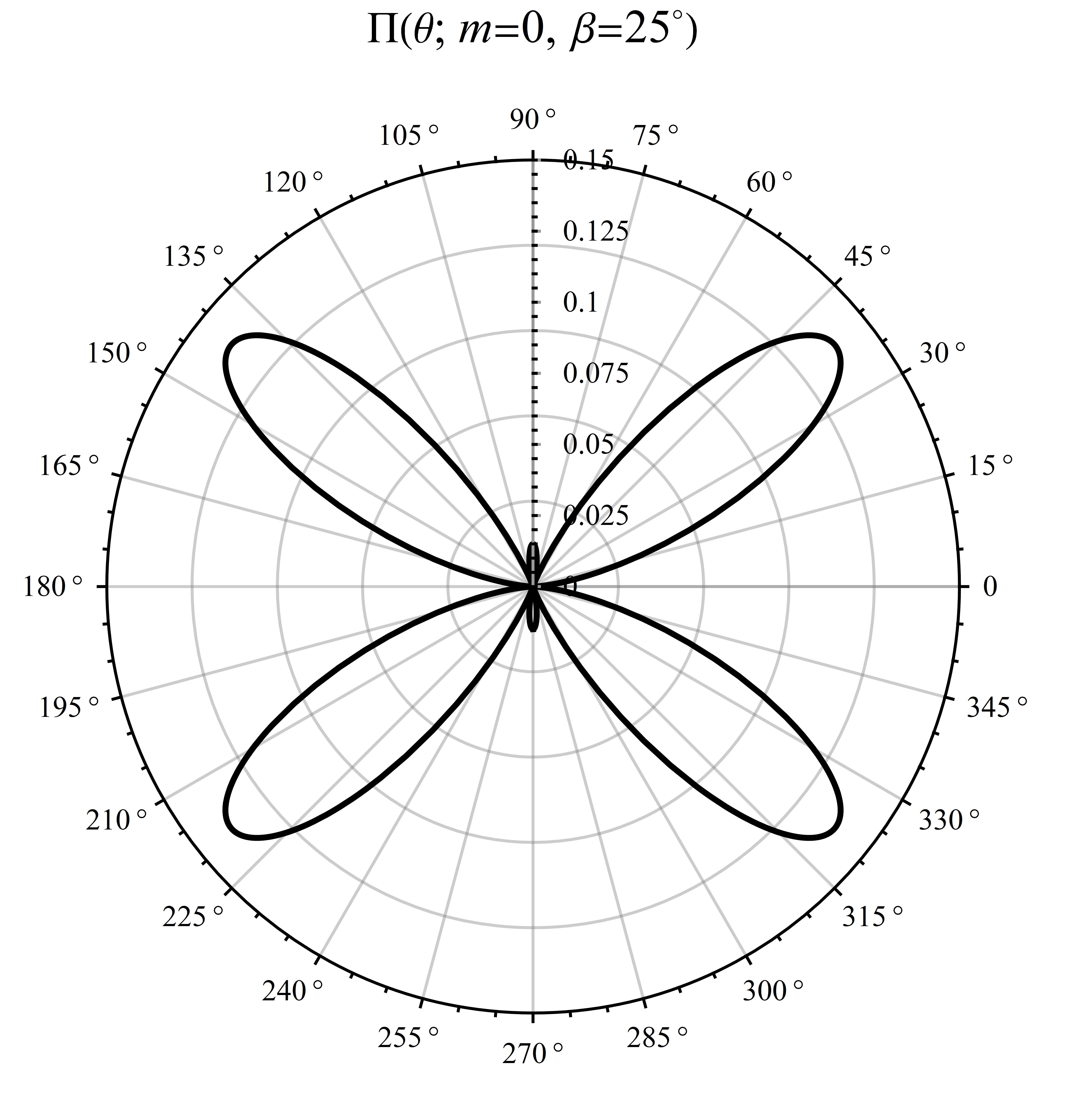}\\
	\includegraphics[scale=0.5]{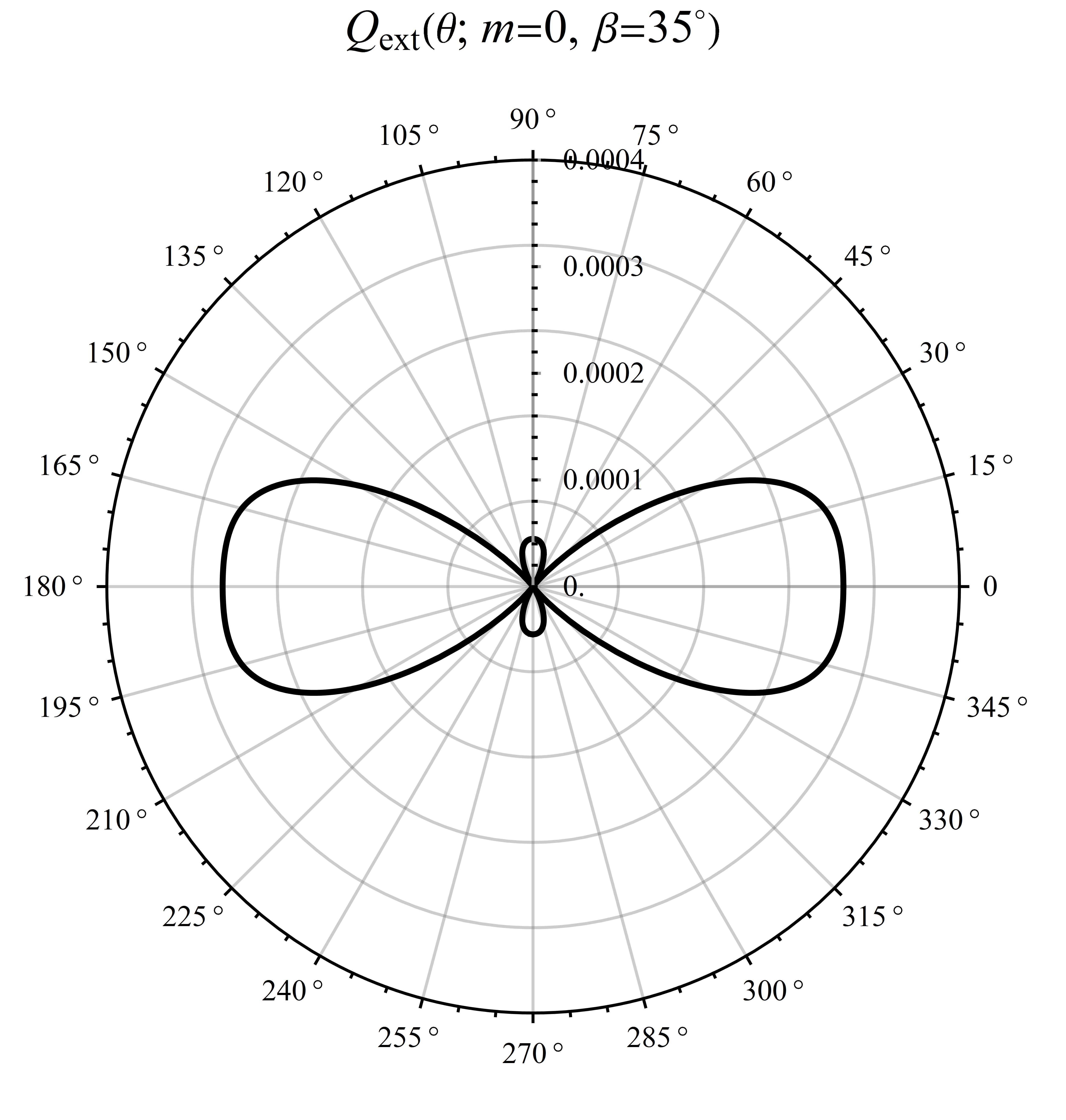}
	\includegraphics[scale=0.5]{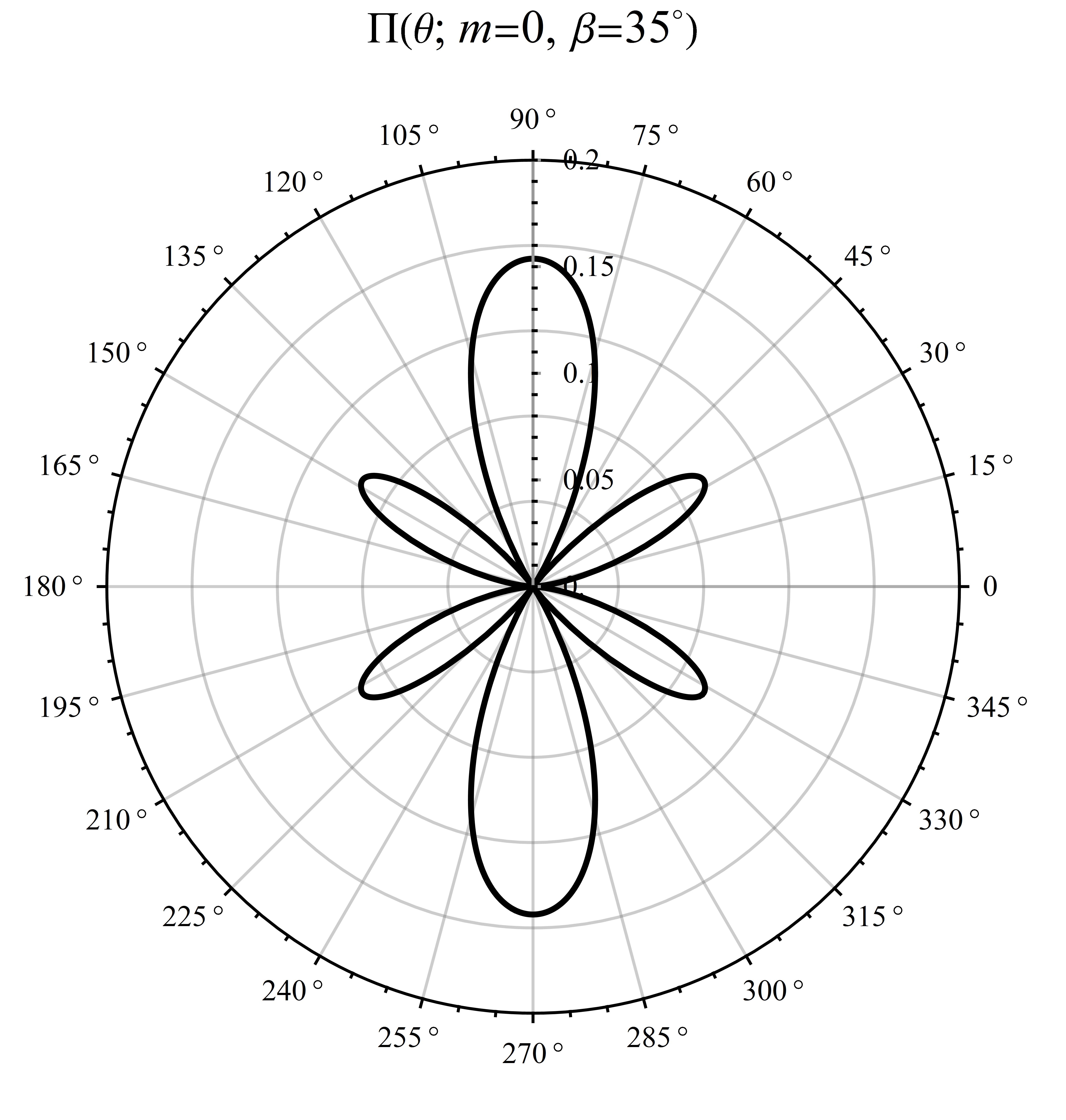}\\
	\includegraphics[scale=0.5]{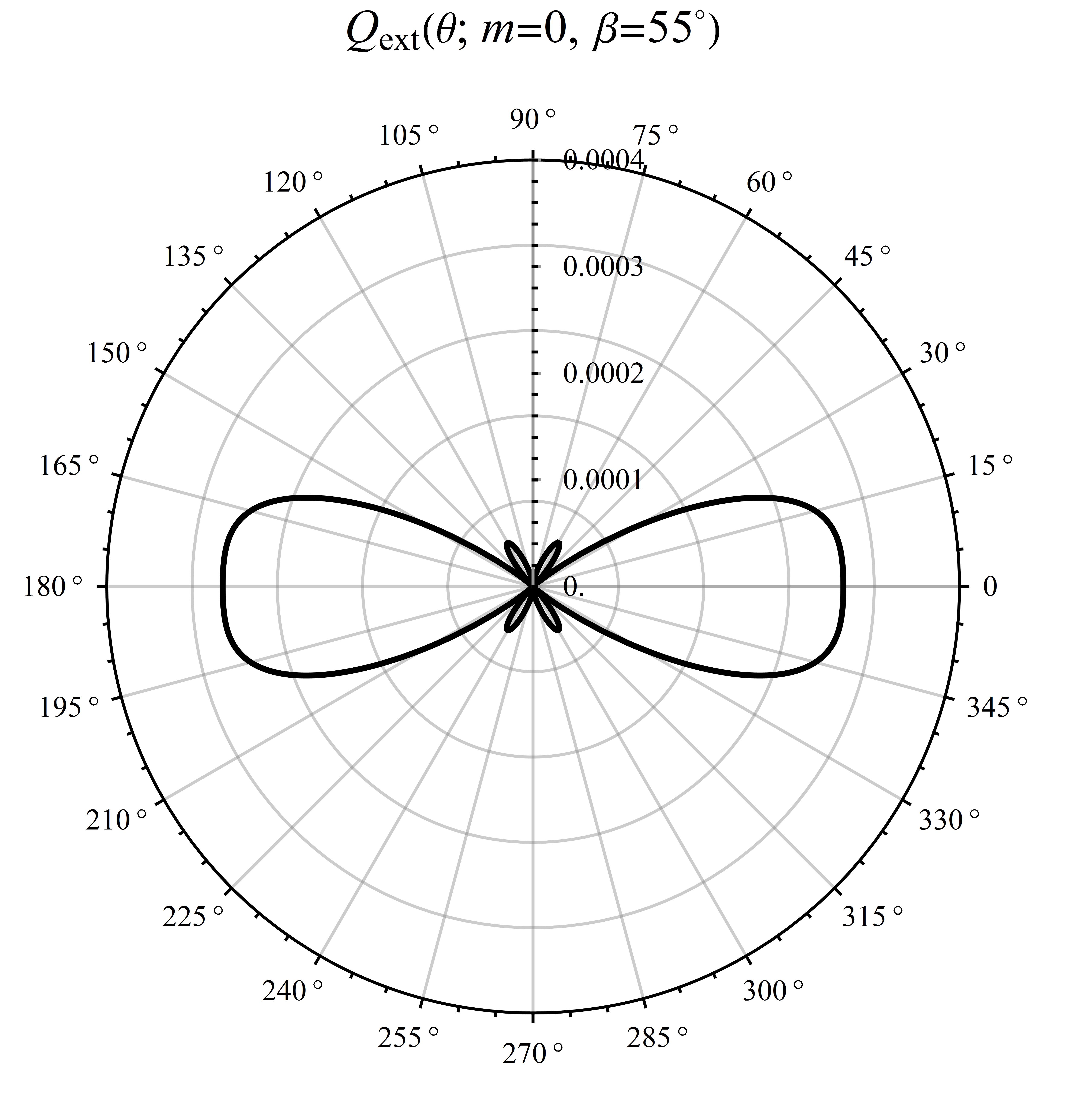}
	\includegraphics[scale=0.5]{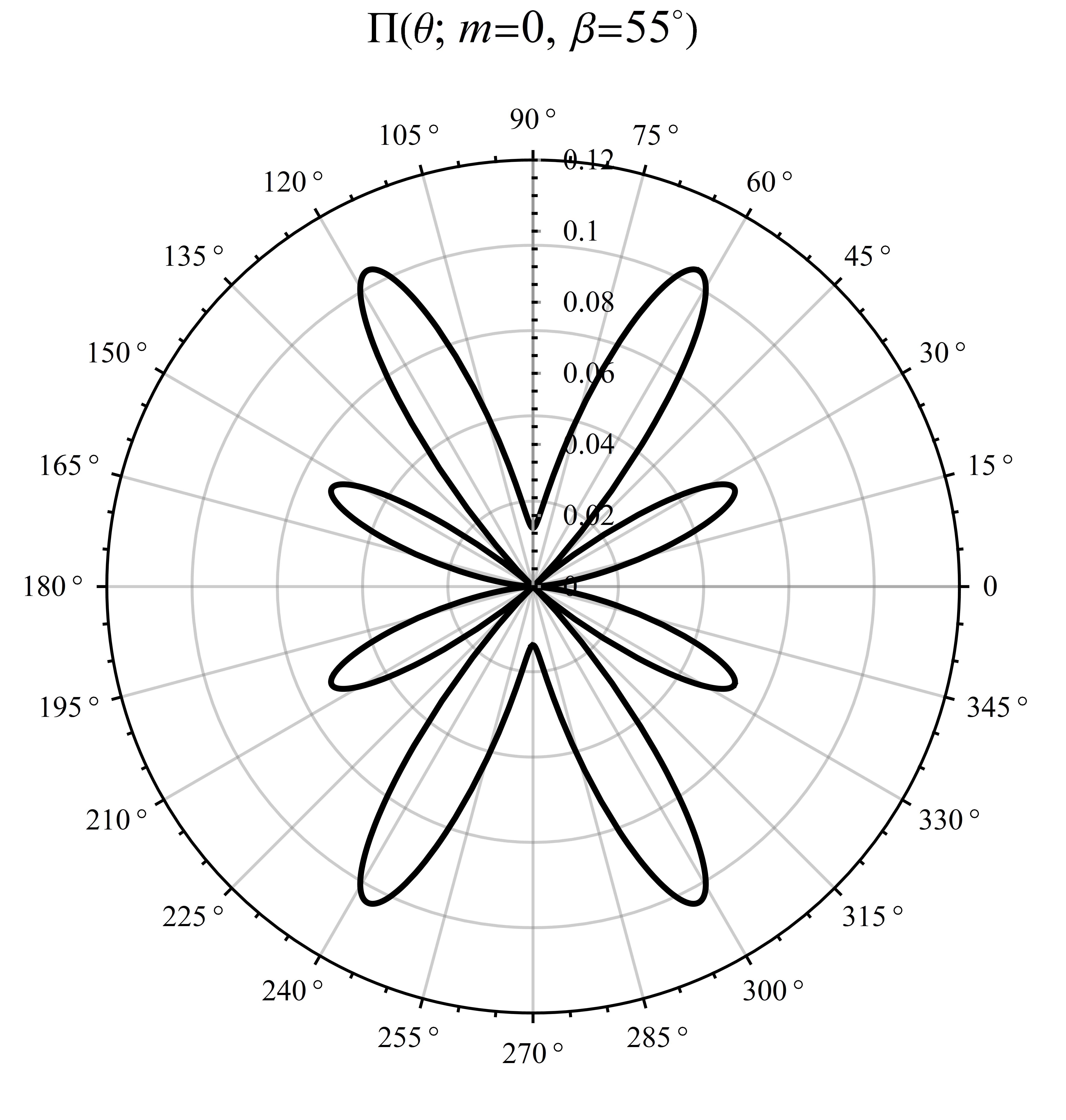}
	\caption{Polar plot of the angular extinction cross section \eqref{ec:Qext}  and the polarization scattered function \eqref{ec:fscatBessel} for a zero order  Bessel beam varying the angle of incidence  to  $\beta= \{ 25^{\circ},  35^{\circ}, 55^{\circ}\}$.}
	\label{figura3}
\end{figure}
Some numerical simulations in which a sphere of radius $ a $, centered at the origin and surrounded by a non absorbing medium, scatters a Bessel beam can be considered at this point; for the sake of simplicity, we will consider only a zero order Bessel beam. Since the zero order Bessel beam has the typical characteristics of a Bessel beam,  it can be easily obtained in the laboratory by different means such as cylindrical lenses (axicon) \cite{Arlt}, a tunable acoustic gradient index of refraction lenses \cite{Leod}, or  using a spatial  light modulator \cite{Victor}. To our knowledge, the Rayleigh scattering problem using this approximation for the invariant beam has not been  addressed  before. However,  the scattering problem of a rigid sphere in acousto-optics using  a zero order  Bessel beam   was firstly  reported in \cite{Marston1,Marston2}; also,  the case of a dielectric sphere has been studied  in \cite{Mitri2011, Zhiwei,HHernandez} for the Mie regime.  \\
Here we look at the analytical scattering  solution  for a dielectric sphere in the Rayleigh regime $ka <<1$, where the transversal vector wave is $k_{T}=  \frac{ 2\pi   }{\lambda}  \sin\theta \sin{\beta}  $  \cite{HHernandez},  being $n$ the index of refraction, $\lambda$ the wavelength and $\beta$ the  cone angle respectively \cite{Durnin1987}. The behavior of the scattering extinction efficiency, given by equation \eqref{ec:Qext}, and the polarization scattering function, represented by equation (36), are shown in Figure \ref{figura3}. We study  particular cases for which the angle of incidence takes the  values  $\beta= 25^{\circ}, 35^{\circ}$  and $55^{\circ}$; in addition, the parameter $ m=0 $, the field wavelength $\lambda=1$ m,  the refractive index  $n =1.5$, the sphere radius $a =0.02$ mm and the adimensional scattering parameter $k = 0.188496$ have been fixed. For  $\beta = 25^{\circ}$,  the total scattering extinction efficiency has two lobes in the forward and backward direction, while four lobes were found for the polarization scattered function. When $\beta = 35^{\circ}$, two additional lobes appear in the total scattering extinction efficiency,   in addition to a forward and backscattering peaks. The former can be interpreted as the peaks of the backscattering and  forward scattering  components of the plane waves constructing the Bessel beam \cite{Marston1,Marston2,Mitri2011}; in this case  the polarization scattered function is zero at $\theta =0^{\circ}$ and   $\theta =180^{\circ}$. Increasing the angle to $ \beta= 55^{\circ}$ narrows the forward and  backscattering peaks, while polarization curves for the forward and backward  scattering  vanish at  $\theta =0^{\circ}$ and $\theta =180^{\circ}$. Similar radiation patterns were found in \cite{Chafiq,Nebdi} for  other members of the family of invariants beams, but it has only been studied for the Mie regime $ka>>1$ and  still  remains unknown for Rayleigh scattering.

\section{Conclusions}
We have obtained a general optical theorem for any propagation invariant beam, where the plane wave case is a particular case. To demonstrate the link between our generalized formulation and prior results in this area, we have shown that the presented ordinary form of the  optical theorem  renders the particular case for free space  derived in previous works. Using the amplitude scattering function in  the far field approximation we have obtained a general representation for  the Rayleigh scattering regime that can be applied to  any invariant beam.  We have studied the scattering of  a Bessel beam, assuming that  the incident wave is linearly polarized and considering  a zero Bessel beam  as a function of  the incident impinging angle $\beta$. Our  method can be  extended to other propagation  invariant beam members, both   to calculate the scattering amplitude function and  to evaluate the forward-scattering approach. This method can be applied  to arbitrary probing fields as  any invariant beam can be written as a plane wave superposition.

\section{Acknowledgments}
The authors are grateful to  B.M. Rodr\'iguez-Lara for his continuous support and for  fruitful  discussions.  I. Rond\'on-Ojeda acknowledges financial support through  CONACyT doctoral grant CVU $423320$.

\end{document}